\documentclass[11pt]{article}

\usepackage{cite}
 \usepackage{subfigure}
\usepackage{multirow}
\usepackage{helvet}
\usepackage{amsmath}
\usepackage{amssymb}
\usepackage{setspace}
\usepackage{graphicx}
\usepackage{empheq}
\usepackage{soul}
\usepackage{tabularx}
\usepackage{array}
\usepackage{float}
\usepackage{braket}
\usepackage[utf8]{inputenc}
\usepackage{url}
\usepackage{hyperref}
\usepackage{slashed}
\usepackage[font=footnotesize,labelfont=bf]{caption}
\usepackage{color}

\def\be{\begin{equation}}
\def\ee{\end{equation}}

\usepackage[a4paper,top=3cm,bottom=3cm,left=2.5cm,right=2.5cm,bindingoffset=0mm]{geometry}

\begin{document}

\begin{center}
{\Large \bf A new calculation of semi--exclusive axion--like particle\\ \vspace*{0.2cm}  production at the LHC}

\vspace*{1cm}

{\sc 
L.A.~Harland-Lang\footnote[1]{email: l.harland-lang@ucl.ac.uk}%
, M.~Tasevsky\footnote[2]{email: Marek.Tasevsky@cern.ch}%
}
\vspace*{0.5cm}

{\small\sl
  $^1$Rudolf Peierls Centre, Beecroft Building, Parks Road, Oxford, OX1 3PU, UK

\vspace*{0.25cm} 

$^2$Institute of Physics of the Czech Academy of Sciences, 
Na Slovance 1999/2, \\
18221 Prague 8, Czech Republic

}

\begin{abstract}
\noindent We present a new calculation of axion--like particle (ALP) production in the semi–exclusive photon--initiated (PI) channel, that is either with intact outgoing protons or rapidity gaps present in the final state, and with no colour flow between the colliding LHC protons. This is the first full treatment of this process, accounting for both the possibility of proton dissociation and  the survival factor probability of no additional proton--proton interactions, including its kinematic and process dependence. We present the expected event yields in the case that either one or both protons are required to be registered in the forward proton detectors installed in association with ATLAS and CMS. We find that this process can be sensitive to currently unprobed regions of ALP mass and ALP--photon coupling, in particular in the intermediate  mass region. Our calculation is provided in the publicly available  \texttt{SuperChic 4} Monte Carlo (MC) generator, and can be passed to a general purpose MC for showering and hadronization of the final state.

\end{abstract}

\end{center}

\section{Introduction}

The existence of new axion--like particles (ALPs) is widespread in extensions of the Standard Model (SM)~\cite{Ringwald:2014vqa}. While the name derives from the well known QCD axion~\cite{Peccei:1977hh,Peccei:1977ur,Wilczek:1977pj,Weinberg:1977ma} these light, gauge--singlet pseudoscalar particles occur in a broad spectrum of beyond the Standard Model (BSM) theories. They occur generically in  theories with a spontaneously broken global symmetries as well as in the low energy effective theory of string compactifications~\cite{Svrcek:2006yi,Arvanitaki:2009fg}. As such, these particles have been the focus of broad ranging and dedicated theoretical and experimental investigations which cover a wide range of ALP mass ranges, from the sub--MeV to the TeV range, see e.g.~\cite{Essig:2013lka,Mimasu:2014nea,Graham:2015ouw,Jaeckel:2015jla,Bauer:2017ris,Irastorza:2018dyq,Fortin:2021cog,dEnterria:2021ljz} for an overview.

These ALPs are often naturally coupled to the electroweak (EW) sector of the SM, with in particular the two--photon production and decay mechanism playing an important role. With this in mind, it was shown in~\cite{Knapen:2016moh} that the $\gamma\gamma$ final state in ultraperipheral heavy ion collisions at the LHC can be used as an effective search channel for such ALPs in the GeV region. Indeed subsequently to this, in~\cite{Knapen:2017ebd} the ATLAS evidence for light--by--light scattering~\cite{ATLAS:2017fur} in ultraperipheral $PbPb$ collisions, was recast to place limits on the existence of such ALPs in the 10--100 GeV mass region (see~\cite{ATLAS:2019azn} for the first observation). More recently, further bounds have been placed by both the ATLAS~\cite{ATLAS:2020hii} and CMS~\cite{CMS:2018erd} collaborations, in the same channel. Indeed, more broadly ultraperipheral heavy ion collisions represent a particularly promising avenue to probe BSM physics with EW couplings, with recent studies of the $\tau$ lepton anomalous magnetic moment being a particularly topical example~\cite{Beresford:2019gww,Dyndal:2020yen,Jofrehei:2022bwh}, see also~\cite{Bruce:2018yzs} for a general review.

On the other hand, as discussed in~\cite{Bruce:2018yzs}, heavy ion collisions have the disadvantage that the initial--state photon flux, though enhanced by $\sim Z^2$, falls significantly with the invariant mass of the produced object. For this reason it has limited reach to the region of higher masses, beyond the $\sim 50 $ GeV region, with the precise value depending on the  couplings assumed and experimental conditions. Above these masses, LHC constraints exist from $\gamma\gamma$ production in both fully inclusive and VBF topologies, see~\cite{Bauer:2018uxu,Florez:2021zoo,dEnterria:2021ljz} for more recent studies. In this region, a promising and complementary alternative is to instead consider exclusive production in $pp$ collisions, as first discussed in~\cite{Baldenegro:2018hng}, where it was shown that this channel could place competitive constraints on ALPs with masses up to the TeV scale (see also~\cite{Shao:2022cly} for a recent Monte Carlo implementation of exclusive photon--initiated production, including of ALPs). In this process, the ALPs are produced in the photon--initiated (PI) channel, with the protons remaining intact following the collision. These can then be measured by the dedicated AFP~\cite{AFP,Tasevsky:2015xya} and CT--PPS~\cite{CT-PPS}  forward proton detectors (FPDs), which have been installed in association with both ATLAS and CMS, respectively. Indeed, searches for exclusive diphoton production in $pp$ collisions have been reported by CT--PPS in~\cite{CMS:2020rzi,CMS:2022zfd}, with the corresponding limits on ALPs reported in~\cite{CMS:2022zfd}.

While the study of~\cite{Baldenegro:2018hng} considered the case that both protons remain intact and are tagged by the FPDs, we can in principle loosen this requirement while still maintaining the potential for probing ALPs at the LHC. Indeed,  the PI channel can in general be probed experimentally by simply requiring that the relevant object of interest be produced in the presence of rapidity gaps, either alone or in the presence of a single proton tag in the FPD. More commonly, during regular LHC running there will be non--zero pileup, and hence one requires that no additional tracks associated with the primary vertex be present; in this paper we will consider event yields corresponding to this scenario. This approach has been taken for the case of both lepton and more recently $W$ boson pair production at the LHC~\cite{Aad:2015bwa,Aaboud:2016dkv,Aaboud:2017oiq,Chatrchyan:2011ci,CMS:2013hdf,Khachatryan:2016mud,Cms:2018het}. The efficiency of this track veto depends on the width of the window around primary vertex, track $p_{\perp}$, and amount of pile-up but was shown in the cited works to be decent for the pile-up levels accessible in Run~1 and 2. Of particular note here is the first measurement of semi--exclusive lepton pair production with a single proton tag~\cite{Cms:2018het,ATLAS:2020mve}.

The key element in the above cases is that by requiring the presence of a single proton tag and/or rapidity gaps in the central detector one can effectively select for the topology of no colour flow between the colliding proton beams. This is naturally the case in PI production, due to the colour--singlet nature of the photon, and hence the PI signal is enhanced. On other hand, without tagging both protons we will have that semi--exclusive production contributes automatically to the signal, that is where one of the protons emit a photon inelastically.

In this paper we present a theoretical treatment of both exclusive and semi--exclusive ALP production in the PI channel, relevant for future LHC searches. This is based on the approach outlined in~\cite{Harland-Lang:2019eai,Harland-Lang:2021zvr,Harland-Lang:2020veo,Bailey:2022wqy}, namely the structure function (SF) approach discussed in~\cite{Harland-Lang:2019eai,Harland-Lang:2021zvr} is used to model the underlying PI process, which is then combined with a fully differential modelling of the survival factor probability of no additional particle production due to multi--parton interactions (MPI). This has been implemented in the  \texttt{SuperChic 4} Monte Carlo (MC) generator~\cite{SuperCHIC} in a form that could then subsequently be passed to a general purpose MC such as \texttt{Pythia}~\cite{Sjostrand:2014zea} for showering and hadronization of the proton dissociation products.

Using this framework, we can then provide accurate predictions for the expected yields from purely elastic (EL) production, as well as single dissociation (SD) and double dissociation (DD), where one or both protons break up, respectively. A key question, discussed recently in~\cite{Bailey:2022wqy}, is to what extent the relative contributions from these components are universal or else process dependent. In contrast to the case of $W^+ W^-$ production considered in this work, we find that for ALP production, in the models without phenomenologically relevant ALP--quark couplings, the relative fractions lie very close to the case of lepton pair production. Thus the case of lepton pair production can to good approximation be used as a proxy to predict the contribution from dissociation production with respect to the purely elastic case, in the PI channel. Nonetheless, this matching is not exact and the MC implementation presented is the more precise one.

We therefore consider the expected event yields for single and double tagged ALP production, relevant to both the AFP~\cite{AFP,Tasevsky:2015xya} and CT--PPS~\cite{CT-PPS} FPDs. We consider a range of integrated luminosities, from $10-300$ ${\rm fb}^{-1}$, that is spanning the region from the current level of accumulated data with tagged protons, to the final expected yield from the LHC Run 3. We find that the expected Run 3 yields in the higher mass, $m_a>150$  GeV, region are at the borderline of the excluded region of parameter space that one gets from suitably interpreting the ATLAS resonance searches~\cite{ATLAS:2017ayi,ATLAS:2021uiz} in the inclusive $\gamma\gamma$ channel, as is first evaluated in~\cite{Bauer:2018uxu}. We find that is particularly true in the double tagged case, where the expected yield in this region of parameter space amounts to a handful of events. However, it is worth emphasising, as shown in~\cite{Baldenegro:2018hng} for the double tag case, that the expected backgrounds are very low, significantly so in contrast to the inclusive channel; the latter point is emphasised in e.g.~\cite{dEnterria:2021ljz}. We in addition consider the impact of requiring a veto on additional tracks. While this is only possible in the case when the final--state photons convert into leptons, we can expect a reasonable fraction of signal events to satisfy this, and in this case such a veto may help to reduce backgrounds. However, it is worth emphasising that such a cut is not expected to be essential once a proton is required; the dominant combinatorial background from pile--up events in association with inclusive diphoton production can be controlled by data--driven means, for example using event mixing, see e.g.~\cite{CMS:2022zfd}.

In the lower mass region, $ 80 \lesssim m_a < 150$ GeV (where the lower limit is set by the relevant experimental $E_\perp^\gamma$ cuts), we find there exist regions of parameter space that are not excluded by current limits and for which the single tagged channel exhibits sensitivity, even with a smaller integrated luminosity of $\sim 10$ ${\rm fb}^{-1}$. We find that here, this is currently beyond the reach of the double tagged case, due the kinematic acceptance in this case. We therefore expect that semi--exclusive ALP production, selected by a single proton tag, can provide not just a complementary constraint on the higher mass ALP region but potentially extend existing searches in the lower mass region at the LHC. The current paper provides the necessary MC simulation to achieve this.

The outline of this paper is as follows. In Section~\ref{sec:model} we outline the approach we use to modelling semi--exclusive ALP production and describe the implementation in \texttt{SuperChic 4}. In Section~\ref{sec:results} we explore the predictions of this approach for single and double tagged (semi)--exclusive ALP production in the PI channel, and comment on our results in light of future LHC searches. Finally, in Section~\ref{sec:conc} we conclude.

\section{Modelling semi--exclusive ALP production}\label{sec:model}

To model ALP production in the exclusive and semi--exclusive channels we  apply the structure function (SF) approach for calculating PI production discussed in~\cite{Harland-Lang:2019eai,Harland-Lang:2021zvr}. In the high--energy limit ($\sqrt{s} \gg m_p$) the PI cross section in proton--proton collisions can be written in the general form
  \be\label{eq:sighh}
  \sigma_{pp} = \frac{1}{2s} \int \frac{{\rm d}^3 p_1' {\rm d}^3 p_2' {\rm d}\Gamma}{E_1' E_2'}   \alpha(Q_1^2)\alpha(Q_2^2)
  \frac{\rho_1^{\mu\mu'}\rho_2^{\nu\nu'} M^*_{\mu'\nu'}M_{\mu\nu}}{Q_1^2 Q_2^2}\delta^{(4)}(q_1+q_2 - k)\;.
 \ee
 Here the outgoing hadronic systems have momenta $p_{1,2}'$ and the photons have momenta $q_{1,2}$, with $q_{1,2}^2 = -Q_{1,2}^2$. The production ALP has 4--momentum $k = q_1 + q_2 = k_1 + k_2 $, where $k_i$ are the outgoing photon momenta, and ${\rm d}\Gamma = \prod_{j=1}^2 {\rm d}^3 k_j / 2 E_j (2\pi)^3$ is the standard phase space volume. 
$\rho$ is the density matrix of the virtual photon, which is given in terms of the well known proton structure functions:
 \be\label{eq:rho}
 \rho_i^{\alpha\beta}=2\int \frac{{\rm d}M_i^2}{Q_i^2}  \bigg[-\left(g^{\alpha\beta}+\frac{q_i^\alpha q_i^\beta}{Q_i^2}\right) F_1(x_{B,i},Q_i^2)+ \frac{(2p_i^\alpha-\frac{q_i^\alpha}{x_{B,i}})(2p_i^\beta-\frac{q_i^\beta}{x_{B,i}})}{Q_i^2}\frac{ x_{B,i} }{2}F_2(x_{B,i},Q_i^2)\bigg]\;,
 \ee
where $x_{B,i} = Q^2_i/(Q_i^2 + M_{i}^2 - m_p^2)$ for a hadronic system of mass $M_i$ and we note that the definition of the photon momentum $q_i$ as outgoing from the hadronic vertex is opposite to the usual DIS convention. Here, the integral over $M_i^2$ is understood as being performed simultaneously with the phase space integral over $p_{i}'$, i.e. is not fully factorized from it (the energy $E_i'$ in particular depends on $M_i$). One can  make use of the wealth of data from lepton--proton scattering to constrain the structure functions, and hence the photon--initiated cross section, to high precision. Full details of the procedure for this are given in~\cite{Harland-Lang:2019eai,Harland-Lang:2021zvr} and are for brevity not repeated here; we note however that the corresponding uncertainties on these inputs is at the 1\% level, and so is under very good control.

In \eqref{eq:sighh} $M^{\mu\nu}$ corresponds to the $\gamma\gamma \to a \to \gamma\gamma$ production amplitude, with arbitrary initial--state photon virtualities. To evaluate this we will consider the following  Lagrangian:
\begin{equation}
\mathcal{L}=\frac{1}{2}\partial^\mu a \partial_\mu a -\frac{1}{2}m_a^2 a^2 -\frac{1}{4}g_a a F^{\mu\nu}\tilde{F}_{\mu\nu}\;,
\end{equation}
where $m_a$ is the ALP mass and $g_a$ is the ALP--photon coupling. As the contribution from $t$ and $u$ channel ALP exchange will be extremely small, and difficult to distinguish from other backgrounds, we will for simplicity consider only the dominant resonant $s$--channel contribution. In this case, the $\gamma (q_1) \gamma (q_2) \to a \to \gamma (k_1) \gamma (k_2)$ amplitude is simply given by
\be\label{eq:alpschan}
M_{\mu\nu} = g_a^2 \epsilon_{\mu\nu \sigma \rho } q_1^\sigma q_2^\rho \left[ \frac{1}{(m_{\gamma\gamma}^2-m_a^2)+ i m_{\gamma\gamma}\Gamma_a}\right ] \epsilon_{\alpha\beta \delta \gamma } \epsilon_1^{\alpha *} \epsilon_2^{\beta *} k_1^\delta k_2^\gamma \;,
\ee
where ALP width is given by
\be\label{eq:alpwidth}
\Gamma_a = \frac{g_a^2 m_a^3}{64\pi}\;,
\ee
under the assumption that the ALP couples only to the photons, i.e. ${\rm Br} (a \to \gamma\gamma)=100\%$, we have
\be
\left \langle M^*_{\mu'\nu'}M_{\mu\nu}\right \rangle \approx g_a^2 \left( \epsilon_{\mu\nu \sigma \rho } q_1^\sigma q_2^\rho  \epsilon_{\mu'\nu' \sigma' \rho' } q_1^{\sigma'} q_2^{\rho'}\right)\left[64\pi^2 \delta(m_{\gamma\gamma}^2-m_a^2)\right ]\;,
\ee
summing over photon polarizations and in the narrow width limit, although in all results we include the full finite width effects as in \eqref{eq:alpschan}. In what follows we will assume that the ALP branching ratio to photons is indeed 100\%, although  one can to good approximation convert to other scenarios straightforwardly, provided the ALP  width remains relatively narrow.

As usual, given the experimental requirement for intact tagged protons and/or a rapidity veto in the central detector, we must evaluate the probability for producing no additional particles produced in the final state. In such a case we need to include the probability of no additional particle production due to soft proton--proton interactions (i.e. underlying event activity), known as the survival factor, see~\cite{Harland-Lang:2014lxa,Harland-Lang:2015eqa} for reviews.
We account for this following exactly the same approach as described in~\cite{Harland-Lang:2020veo}, for the case of lepton pair production, but with the  production amplitude now corresponding to the $\gamma \gamma \to a \to \gamma \gamma$  process described above. This results in a predicted survival factor that depends differentially on the final--state kinematics, as well as the produced object and whether the production is purely elastic or inelastic. As discussed in~\cite{Harland-Lang:2020veo}  we find that the survival factor is rather close to unity for the case of purely elastic production, while being significantly smaller in the double dissociative case.

Using the above formalism, we have implemented ALP production in  \texttt{SuperChic 4}. In particular, events are generated fully differentially in terms of the final--state ALP decay and proton dissociation products, if the process is inelastic. These can then be  interfaced to \texttt{Pythia 8.2}~\cite{Sjostrand:2014zea} for showering and hadronisation of the proton dissociation system. The same approach to achieve this, and corresponding flags in \texttt{Pythia 8.2}, are again exactly as described in~\cite{Harland-Lang:2020veo}, to which we refer the reader for further details.

Finally, we have also implemented in \texttt{SuperChic 4} the case of a scalar ALP coupled to photons, for which the corresponding Lagrangian is
\begin{equation}
\mathcal{L}=\frac{1}{2}\partial^\mu a \partial_\mu a -\frac{1}{2}m_a^2 a^2 -\frac{1}{4}g_a a F^{\mu\nu}F_{\mu\nu}\;,
\end{equation}
and the $\gamma \gamma a$ amplitude becomes
\be
\epsilon_{\mu\nu \sigma \rho } q_1^\sigma q_2^\rho \to  g_{\mu \nu} (q_1 q_2) - q_{1,\mu} q_{2,\nu}\;.
\ee

\section{Results}\label{sec:results}

In this section we show some representative results for ALP production in  LHC $pp$ collisions. We will for concreteness consider the case of a scalar ALP, but note that a pseudoscalar ALP gives rather similar results for the same mass/couplings. More precisely, the predicted survival factor for EL and SD production is $\sim 10\%$ lower in the case of pseudoscalar production, and hence the overall results presented here will be rather similar in that case. We note that the dependence of the survival factor on the parity of the produced ALP is in line with many earlier studies for exclusive production of various states, see e.g.~\cite{Kaidalov:2003fw,Harland-Lang:2010ajr}. We in addition consider the case of purely $s$--channel ALP production: although $t$--channel ALP exchange can readily be included as a contribution to the $\gamma\gamma\to \gamma\gamma$ process (as is done in e.g.~\cite{Baldenegro:2018hng}) we will in general expect this non--resonant contribution to $\gamma\gamma$ production to be rather mild in the allowed region of ALP--$\gamma$ coupling space, and moreover be significantly more challenging to disentangle from the continuum background. We therefore for simplicity do not focus on this here.

We will for concreteness require the produced photons to have $E_\perp > 40$ GeV and $|\eta| <2.37$, i.e. lying in the ATLAS EM calorimeter with sufficient transverse energy, $E_\perp$, to pass standard diphoton L1 triggers and event selection criteria (see e.g.~\cite{ATLAS:2017ayi}). With this, we can consider the mass range  $80 < m_a < 2600$ GeV, with the upper limit driven by the fall off of the event rate within the Run 3 dataset, as we will see. When presenting event yields, we consider a range of integrated luminosities, from $10-300$ ${\rm fb}^{-1}$, that is spanning the region from the current level of accumulated data with tagged protons, to the final expected yield from the LHC Run 3; while we show results for 13 TeV for the sake of comparison, the expected yields will be similar at 14 TeV. We will consider single and double proton tags for two different acceptance scenarios, namely $0.035<\xi_i < 0.08$ and $0.02 < \xi_i < 0.12$, where $\xi_i$ is the fractional momentum loss of proton $i$. While the former corresponds to a rather narrow region used in the ATLAS exclusive dilepton analysis with single--tagged protons~\cite{ATLAS:2020mve}, the latter corresponds to the maximum range accessible in Run~2 by both AFP and CT-PPS. We in addition impose an acoplanarity cut $1-\Delta \phi_{\gamma\gamma}/\pi < 0.01$.  This is necessary in order to suppress backgrounds  from dijets produced in non-diffractive, single-diffractive or central exclusive production, where the photons are present in the set of final state particles or come from misidentifying jets as photons. This cut is less effective in suppressing fake photons from FSR in photon-induced dilepton production or from misidentifying the produced leptons as photons. Loosening this cut increases the SD contribution somewhat, with the impact on the total rate being $\sim 10\%$ or less.  For the case of SD and DD production, it is in general possible that a proton produced from the decay of the beam proton dissociation system can be registered in the acceptance of the FPD (see~\cite{Harland-Lang:2018hmi,Harland-Lang:2020veo} for further discussion). We account for this possibility in our analysis,  requiring that the $\xi$ of the proton registered in the FPD and that reconstructed from the invariant mass and rapidity of the central diphoton system coincide within 0.005 in order to suppress this contribution. We will in addition investigate the effect of imposing a veto on additional tracks with $p_\perp > 0.5$ GeV and $|\eta| <2.5$: during nominal running, and only in the case when the final--state photons convert into leptons (see e.g. Ref.~\cite{ATLAS:2014cnc} for details about the diphoton production vertex determination from two photons and additional tracking information) we can effectively impose such a veto on additional associated tracks in order to suppress the background.

We note that, as discussed in detail in~\cite{Baldenegro:2018hng} in the double tagged case, there are a range of irreducible and reducible SM backgrounds that we must in general consider in both the single and double tagged event selection. The reducible backgrounds due to continuum semi--exclusive $\gamma\gamma$ production (either gluon or photon induced) are expected to be negligible; taking the single tag requirement and the above event selection we find from  \texttt{SuperChic 4} that $\sim 2$ events in total due to purely exclusive production, for $300$ ${\rm fb}^{-1}$, while for the semi--exclusive case a similar number is expected, although a precise calculation of this is not currently available. This is in line with the results of~\cite{Baldenegro:2018hng}, while in that study the double pomeron exchange (DPE) background is also considered. The cross section for this is found to be completely negligible in the relevant kinematic region for double tagged production, while after imposing matching constraints on the reconstructed proton $\xi$ and that of the central diphoton system, which in general do not match for DPE production, this is reduced to effectively zero. For the single tagged case, one can still effectively suppress this background by requiring that the $\xi$ of the one tagged proton matches the relevant kinematics of the central system. 

Finally, there is the effect of additional pile-up collisions to consider, which in general influences the total background in two ways. First,  additional soft tracks and clusters in the central detector deteriorate various reconstruction efficiencies and resolutions, e.g. vertex, track, lepton or photon reconstruction efficiencies. Second,  protons from pile-up collisions may end up in the acceptance of the FPD and form a so-called combinatorial background. The size of this strongly depends on the $\xi$ range of the FPD and on the number of pile-up collisions per bunch crossing, $\mu$, with a simple rule that the narrower $\xi$ range and the lower $\mu$, the less significant this combinatorial background is. Double-tagged events have two advantages over single-tagged events, namely that the appropriate combinatorial factor is the square of that in the case of single-tagged events, and hence the background is correspondingly suppressed.  Moreover, time-of-flight (ToF) detectors can be used to suppress the combinatorial background. On the other hand, as we will see, double tagged events have the disadvantage of  a narrower acceptance in the mass of the central system and significantly lower event yield when compared to single-tagged events. More details can be found in~\cite{Cerny:2020rvp,Tasevsky:2014cpa,Harland-Lang:2018hmi}. It is not the purpose of the current paper to present a detailed analysis of these backgrounds here, but we simply note that the dominant pile-up background should be controllable in Run 2 and 3 conditions by e.g. suitably matching the kinematics of the central system with the tagged proton, although the specific level of rejection will depend on the details of the experimental analysis.
In this context it is important to note that the search for ALP is in fact a bump hunting on the mass distribution of combinatorial background. It is therefore useful to model this background properly rather than trying to reduce it at the cost of reducing the signal statistics, especially in cases when the total integrated luminosity is low. On the other hand, it is worth emphasising that this background is also controllable by data--driven means for example using event mixing, see e.g. ~\cite{CMS:2022zfd}. Moreover, even in the case of a single proton tag one can still match the $\xi$ of the tagged proton to that of the diphoton system in order to suppress this background.

\begin{figure}
\begin{center}
\includegraphics[width=0.495\textwidth,height=6cm]{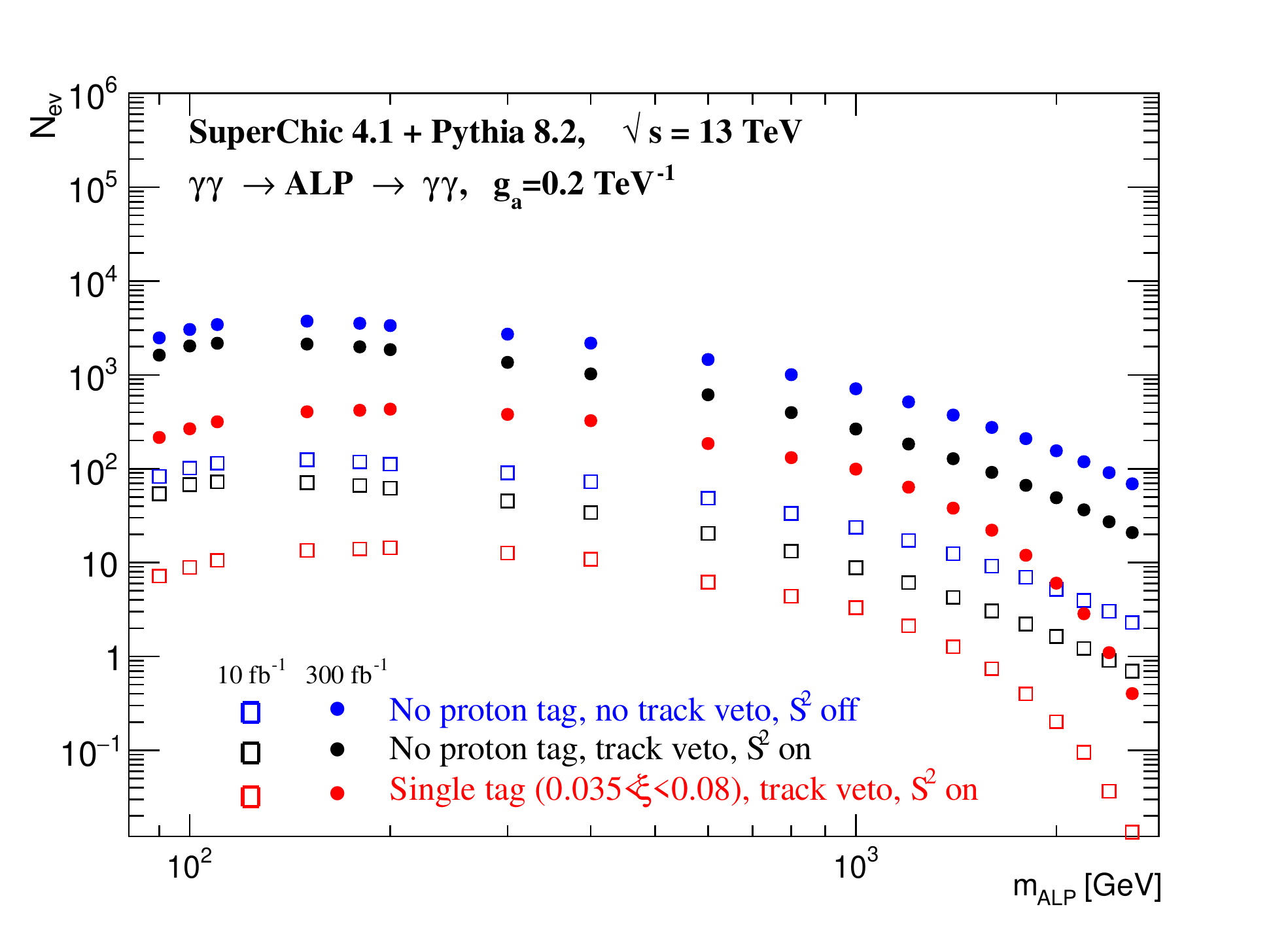}
\includegraphics[width=0.495\textwidth,height=6cm]{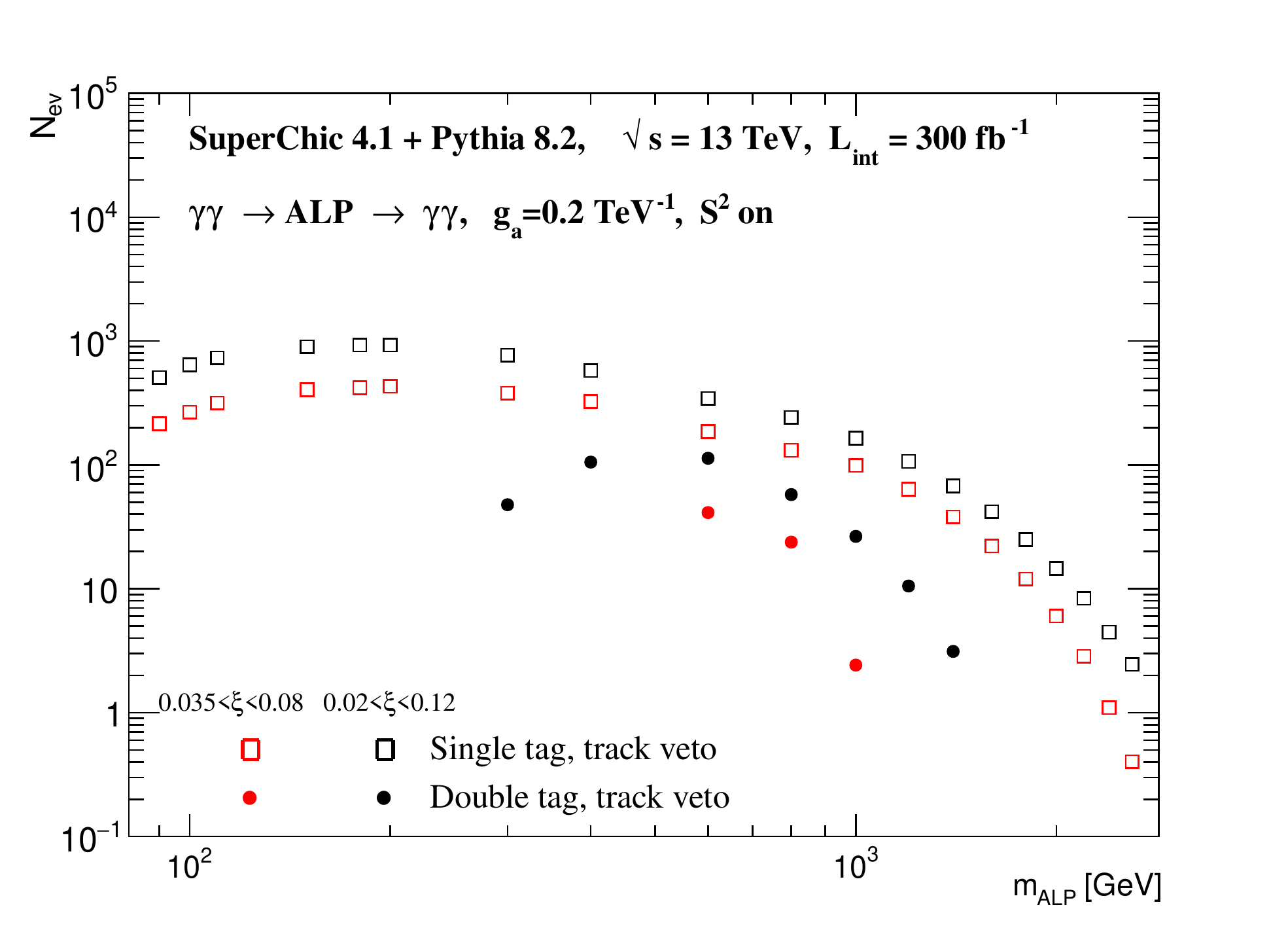}
\caption{Expected event yield for ALP production in 13 TeV $pp$ collisions and for a range of ALP masses, $m_a$. Results shown for a representative coupling $g_a=0.2 \,{\rm TeV}^{-1}$ and assuming ${\rm Br} (a\to \gamma\gamma)=100\%$. (Left) Expected event yields with and without a single proton tag applied, for demonstration, and for $L_{\rm int} = 10-300$ ${\rm fb}^{-1}$. (Right) Expected event yields for single and double proton tags, with $L_{\rm int} = 300$ ${\rm fb}^{-1}$.While a track veto is imposed when indicated, in this single proton tag case, if we remove this then the predicted cross section is only a few percent higher. Note no experimental efficiencies are included in the fiducial cross sections.}
\label{fig:total}
\end{center}
\end{figure}

\begin{figure}
\begin{center}
\includegraphics[scale=0.43]{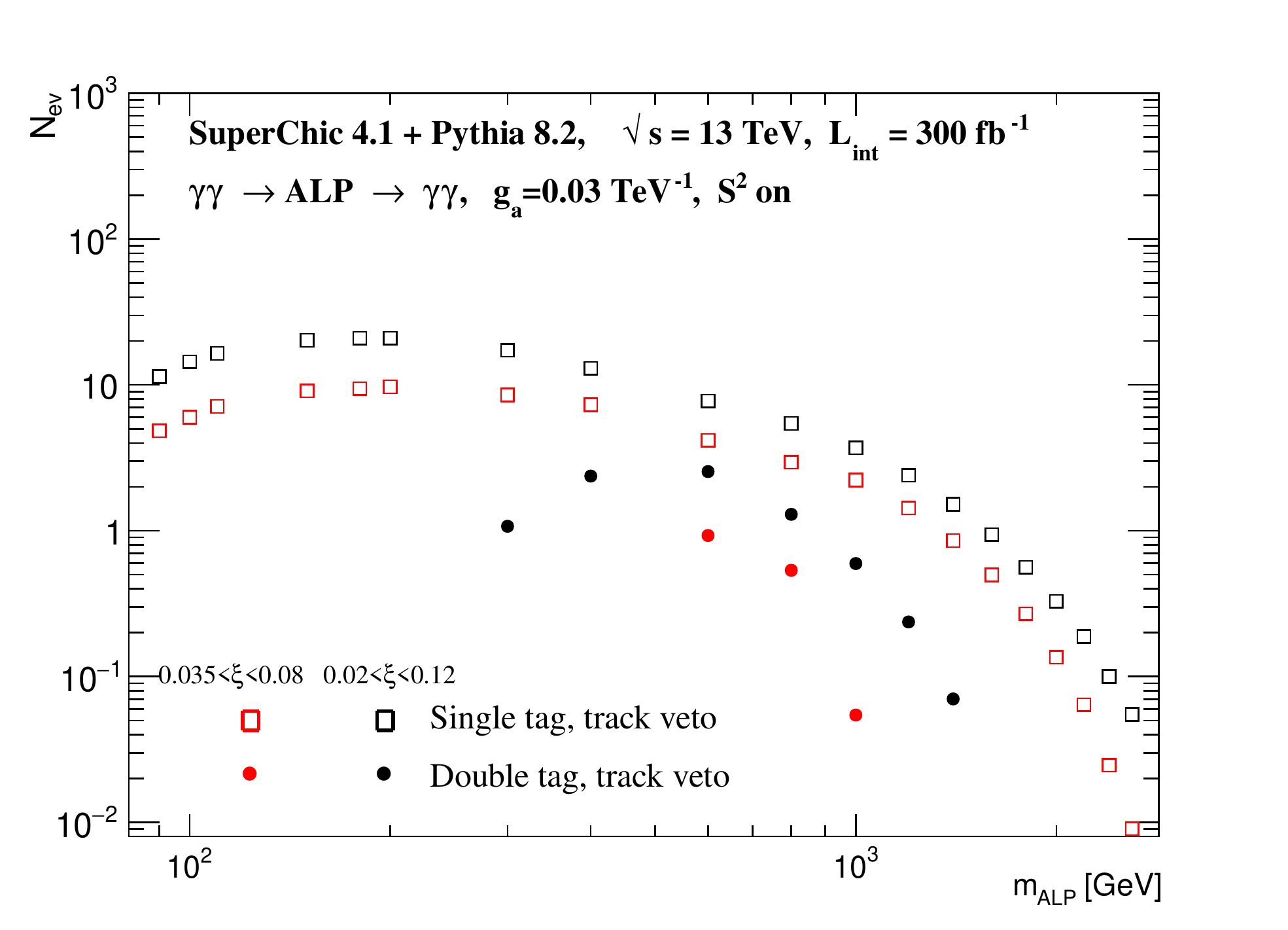}
\caption{As in Fig.~\ref{fig:total} (right) but for $g_a=0.03 \,{\rm TeV}^{-1}$.}
\label{fig:total03}
\end{center}
\end{figure}

With the above discussion in mind, we simply consider in this paper the relevant signal rates, in order to assess whether the single tag search is feasible, and if so in what regions of parameter space, under the reasonable assumption that the relevant backgrounds are under control. In Fig.~\ref{fig:total} (left) we show the expected event yields for 10 and 300 ${\rm fb}^{-1}$, for a range of ALP masses and for  $g_a=0.2 \,{\rm TeV}^{-1}$. We can see from e.g. Fig. 4 (right) of~\cite{dEnterria:2021ljz} and the more recent ATLAS limits in~\cite{ATLAS:2021uiz} that below $m_a < 150$ GeV this region of parameter space is currently not excluded. The limits shown in~\cite{dEnterria:2021ljz} in the 100--160 GeV region correspond to those derived in~\cite{Jaeckel:2012yz}, which come from a reinterpretation of relatively early ATLAS data on Higgs to diphoton production with VBF cuts applied~\cite{Aad:2012tfa}. Although more recent Higgs to diphoton data exist~\cite{CMS:2021kom,ATLAS:2022fnp}, the limits derived in~\cite{Jaeckel:2012yz} are to our knowledge the most recent corresponding limits on ALP production in this intermediate mass region. We can observe that even for 10   ${\rm fb}^{-1}$ of data we expect $\sim 10$ signal events in this region, depending on the precise mass considered. This therefore raises the interesting prospect of extending the LHC into this region of parameter space with existing single proton tag data. In the future, we can see from the plot that a much larger signal yield would naturally be expected. 

We note from \eqref{eq:alpwidth} that the ALP resonance is very narrow in this region of parameter space, and hence to very good approximation the cross section is simply proportional to $g_a^2$, allowing the reader to readily  extract the corresponding signal expectations for other coupling/mass scenarios, provided the width remains narrow in these. Indeed, in Fig.~\ref{fig:total03} we show for illustration the equivalent event yields for a lower coupling $g_a=0.03 \,{\rm TeV}^{-1}$, and we can see that for the larger Run 3 event sample  it may be possible to extend the sensitivity significantly beyond the current exclusion region for  $m_a < 150$ GeV, which extends to $g_a \gtrsim 0.2-0.6$  ${\rm TeV}^{-1}$, depending on the ALP mass. We also show in the plots the expected yields with double proton tags. Due to the kinematic acceptance of this double tag requirement, we can see that a more limited mass range, $260 (455) < m_a < 1560 (1040)$ GeV for the larger (smaller) FPD acceptance range considered, is available. The predicted yields are as expected smaller, due to the additional proton tag, although the control over the corresponding BGs will be improved in this case.

In the  $m_a > 150$ GeV region, on the other hand, the inclusive constraints from~\cite{ATLAS:2017ayi,Bauer:2018uxu} exclude this region, although we note that in~\cite{dEnterria:2021ljz} it is suggested that the treatment of the corresponding backgrounds  in this case may be too optimistic. At larger $m_a>600$ GeV, the recent CT--PPS search~\cite{CMS:2022zfd} with tagged protons sets similar, though slightly less constraining, limits. Nonetheless, even in this region of parameter space we may consider the semi--exclusive ALP production as a complementary cross check of the corresponding inclusive exclusions limits in a potentially cleaner channel to the inclusive case. We note that the value of the coupling taken in Fig.~\ref{fig:total03} is on the edge of or below the current exclusions from these inclusive constraints, and hence there is the potential for extending the sensitivity to somewhat lower couplings even for $m_a > 150$ GeV, with the full Run 3 data sample.

\begin{figure}
\begin{center}
\includegraphics[width=0.49\textwidth,height=6cm]{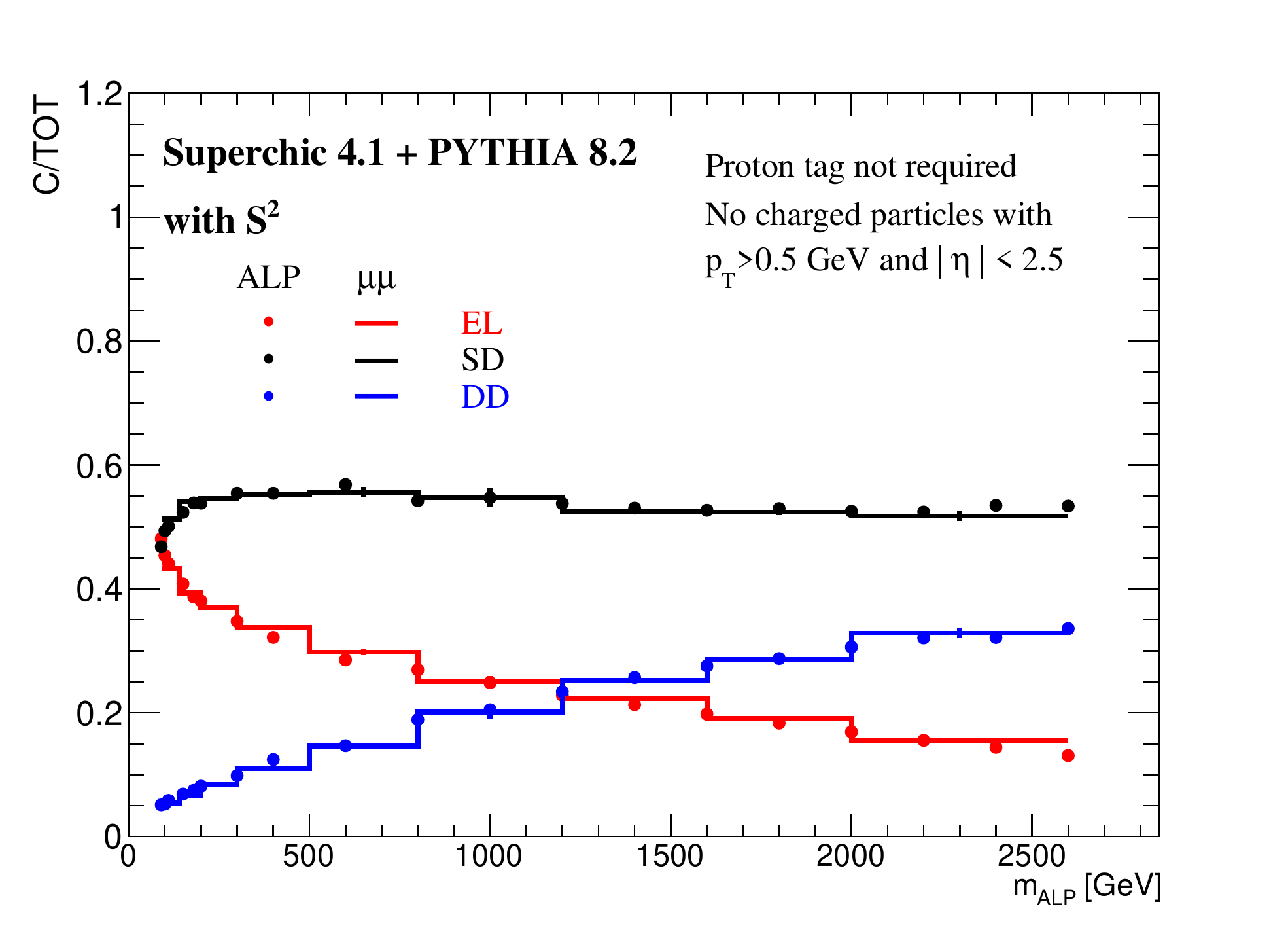}
\includegraphics[width=0.45\textwidth,height=6cm]{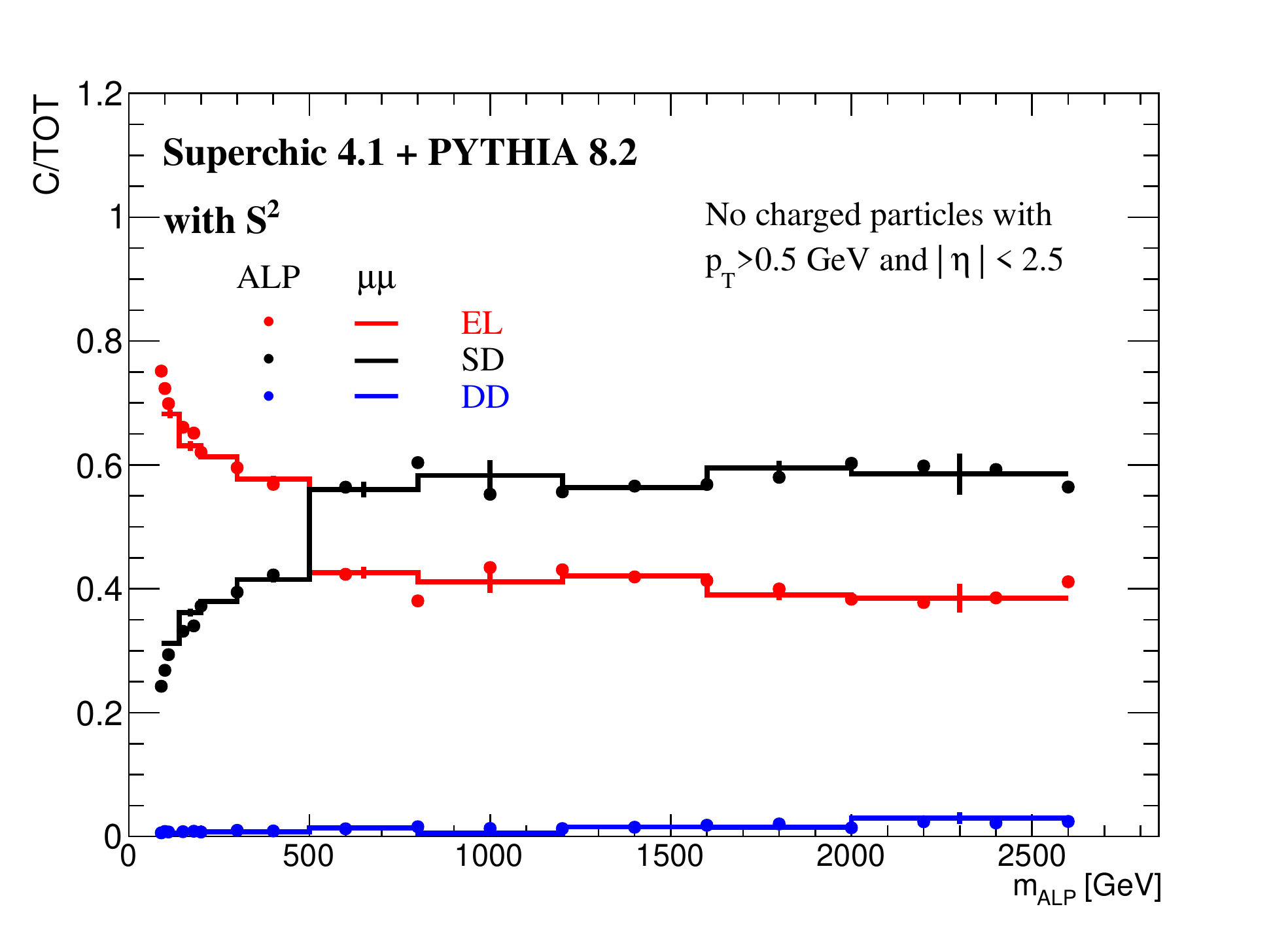}
\caption{Predicted breakdown between elastic, single and double dissociative production for ALP production, and for a range of ALP masses, $m_a$. The left (right) plots shows the result without (with) the additional single proton tag in the FPD acceptance $0.035 < \xi < 0.08$.  Results shown for a representative coupling $g_a=0.2 \,{\rm TeV}^{-1}$ and assuming ${\rm Br} (a\to \gamma\gamma)=100\%$. A track veto is imposed in both plots, though for the single proton tag case (right), if we remove this then the predicted cross section is only a few percent higher.}
\label{fig:breakdown}
\end{center}
\end{figure}

In Fig.~\ref{fig:total} (left) we also show for comparison the predicted yields without a single proton tag, and without a veto imposed. We can see as expected that the single tag reduces the final rate in a way that depends on the mass of the ALP. We note that the dominant effect here is the proton tag itself; if we ask for a single proton tag but do not impose a track veto in the central detector region then the predicted cross section is only a few percent higher. In Fig.~\ref{fig:breakdown} we show the breakdown between the three cross section components, namely the EL, SD and DD. In the left plot we show results without the single proton tag imposed, and as expected the total cross section is a combination of all three components, with the precise amounts depending on the mass of the produced system (similar effects are seen in~\cite{Bailey:2022wqy,Harland-Lang:2020veo}). Once the proton tag is required, however, the DD contribution is negligible, while the EL and SD become relatively enhanced. We note that the DD is not entirely zero due to the effect discussed above, namely that a proton produced from the dissociation system may still register in the FPD. The change in behaviour of the EL and SD rates around $m_a \approx 500$ GeV is due to the fact below $m_a < \xi_{\rm min} \sqrt{s}$, there is no acceptance for ALP production at central rapidity, and hence the ALP must be produced at either forward or backward rapidity. This leads to a non--trivial dependence in the FPD acceptance that behave differently in the EL case, where we require that either of the EL protons registers in the FPD, and SD case, where there is only one elastic proton to register. Finally, we also compare our predictions with the expectation for lepton pair production at similar mass intervals. The agreement is found to be very good, due to the fact that in both cases the pure PI channel is dominant. This is in contrast to the case of $W^+ W^-$ production, as discussed in~\cite{Harland-Lang:2020veo}, and indicates that indeed lepton pair production can act as a useful proxy to evaluate the relative elastic and inelastic components of the ALP production cross section. The agreement is not exact however, and the more precise treatment presented here is to be preferred.

\section{Conclusions}\label{sec:conc}

In this paper, we have presented the first full theoretical description of semi--exclusive photon--initiated (PI) axion--like particle (ALP) production in proton--proton collisions, that is where the outgoing protons either remain intact or dissociate, but with no colour flow between the colliding beams. This accounts for the full kinematics of the production process and the survival factor probability, as well as providing an implementation in the  \texttt{SuperChic 4} MC, which can then be passed to a general purpose detector for showering/hadronization of the proton dissociation system. The code and a user manual for which can be found at 
\\
\\
\href{http://projects.hepforge.org/superchic}{{\tt http://projects.hepforge.org/superchic}}
\\
\\ 
This process is of particular interest as a search channel for relatively high mass ALPs decaying to two photons, with either a single or double proton tag in the FPDs associated with ATLAS and CMS allowing for an effective event selection and suppression of the corresponding backgrounds. In the single tag case, a full treatment of the semi--exclusive mode, where one or both protons dissociate is essential, as has been presented in this work.

We have presented expected  signal yields for both the single and double tagged scenarios, accounting for the experimentally relevant event selection, and considered the basic implications of these, in light of the current experimental constraints from the LHC and elsewhere. We find that while the higher mass, $m_a > 150$ GeV, region of parameter space that can be probed in this channel is already well constrained by the corresponding  PI production contribution to inclusive resonance searches, there are still possibilities currently to extend this with the full Run 3 dataset. More broadly this provides a complementary confirmation of the results in a channel where the backgrounds are potentially lower. At lower masses, $m_a < 150$ GeV, there exist regions of unconstrained parameter space that may be probed even with a relatively small dataset. Semi--exclusive ALP production therefore represents a  promising search channel for ALPs, and in this paper we have presented the theoretical and MC tools necessary to achieve this.

\section*{Acknowledgements}

We thank Gen Tateno, Cristian Baldenegro Barrera, David d'Enterria, Valery Khoze and Misha Ryskin for useful discussions. LHL thanks the Science and Technology Facilities Council (STFC) for support via grant award ST/L000377/1. MT is supported by MEYS of the Czech Republic within project LTT17018.

\bibliography{references}{}

\begin{thebibliography}{10}

\bibitem{Ringwald:2014vqa}
A.~Ringwald,
\newblock {Axions and Axion-Like Particles},
\newblock in {\em {49th Rencontres de Moriond on Electroweak Interactions and
  Unified Theories}}, pp. 223--230, 2014, 1407.0546.

\bibitem{Peccei:1977hh}
R.~D. Peccei and H.~R. Quinn,
\newblock Phys. Rev. Lett. {\bf 38}, 1440 (1977).

\bibitem{Peccei:1977ur}
R.~D. Peccei and H.~R. Quinn,
\newblock Phys. Rev. D {\bf 16}, 1791 (1977).

\bibitem{Wilczek:1977pj}
F.~Wilczek,
\newblock Phys. Rev. Lett. {\bf 40}, 279 (1978).

\bibitem{Weinberg:1977ma}
S.~Weinberg,
\newblock Phys. Rev. Lett. {\bf 40}, 223 (1978).

\bibitem{Svrcek:2006yi}
P.~Svrcek and E.~Witten,
\newblock JHEP {\bf 06}, 051 (2006), hep-th/0605206.

\bibitem{Arvanitaki:2009fg}
A.~Arvanitaki, S.~Dimopoulos, S.~Dubovsky, N.~Kaloper, and J.~March-Russell,
\newblock Phys. Rev. D {\bf 81}, 123530 (2010), 0905.4720.

\bibitem{Essig:2013lka}
R.~Essig {\em et~al.},
\newblock {Working Group Report: New Light Weakly Coupled Particles},
\newblock in {\em {Community Summer Study 2013}: {Snowmass on the
  Mississippi}}, 2013, 1311.0029.

\bibitem{Mimasu:2014nea}
K.~Mimasu and V.~Sanz,
\newblock JHEP {\bf 06}, 173 (2015), 1409.4792.

\bibitem{Graham:2015ouw}
P.~W. Graham, I.~G. Irastorza, S.~K. Lamoreaux, A.~Lindner, and K.~A. van
  Bibber,
\newblock Ann. Rev. Nucl. Part. Sci. {\bf 65}, 485 (2015), 1602.00039.

\bibitem{Jaeckel:2015jla}
J.~Jaeckel and M.~Spannowsky,
\newblock Phys. Lett. B {\bf 753}, 482 (2016), 1509.00476.

\bibitem{Bauer:2017ris}
M.~Bauer, M.~Neubert, and A.~Thamm,
\newblock JHEP {\bf 12}, 044 (2017), 1708.00443.

\bibitem{Irastorza:2018dyq}
I.~G. Irastorza and J.~Redondo,
\newblock Prog. Part. Nucl. Phys. {\bf 102}, 89 (2018), 1801.08127.

\bibitem{Fortin:2021cog}
J.-F. Fortin {\em et~al.},
\newblock Int. J. Mod. Phys. D {\bf 30}, 2130002 (2021), 2102.12503.

\bibitem{dEnterria:2021ljz}
D.~d'Enterria,
\newblock {Collider constraints on axion-like particles},
\newblock in {\em {Workshop on Feebly Interacting Particles}}, 2021,
  2102.08971.

\bibitem{Knapen:2016moh}
S.~Knapen, T.~Lin, H.~K. Lou, and T.~Melia,
\newblock Phys. Rev. Lett. {\bf 118}, 171801 (2017), 1607.06083.

\bibitem{Knapen:2017ebd}
S.~Knapen, T.~Lin, H.~K. Lou, and T.~Melia,
\newblock CERN Proc. {\bf 1}, 65 (2018), 1709.07110.

\bibitem{ATLAS:2017fur}
ATLAS, M.~Aaboud {\em et~al.},
\newblock Nature Phys. {\bf 13}, 852 (2017), 1702.01625.

\bibitem{ATLAS:2019azn}
ATLAS, G.~Aad {\em et~al.},
\newblock Phys. Rev. Lett. {\bf 123}, 052001 (2019), 1904.03536.

\bibitem{ATLAS:2020hii}
ATLAS, G.~Aad {\em et~al.},
\newblock JHEP {\bf 11}, 050 (2021), 2008.05355.

\bibitem{CMS:2018erd}
CMS, A.~M. Sirunyan {\em et~al.},
\newblock Phys. Lett. B {\bf 797}, 134826 (2019), 1810.04602.

\bibitem{Beresford:2019gww}
L.~Beresford and J.~Liu,
\newblock Phys. Rev. D {\bf 102}, 113008 (2020), 1908.05180.

\bibitem{Dyndal:2020yen}
M.~Dyndal, M.~Klusek-Gawenda, M.~Schott, and A.~Szczurek,
\newblock Phys. Lett. B {\bf 809}, 135682 (2020), 2002.05503.

\bibitem{Jofrehei:2022bwh}
CMS, A.~Jofrehei,
\newblock (2022), 2205.05312.

\bibitem{Bruce:2018yzs}
R.~Bruce {\em et~al.},
\newblock J. Phys. G {\bf 47}, 060501 (2020), 1812.07688.

\bibitem{Bauer:2018uxu}
M.~Bauer, M.~Heiles, M.~Neubert, and A.~Thamm,
\newblock Eur. Phys. J. C {\bf 79}, 74 (2019), 1808.10323.

\bibitem{Florez:2021zoo}
A.~Fl\'orez {\em et~al.},
\newblock Phys. Rev. D {\bf 103}, 095001 (2021), 2101.11119.

\bibitem{Baldenegro:2018hng}
C.~Baldenegro, S.~Fichet, G.~von Gersdorff, and C.~Royon,
\newblock JHEP {\bf 06}, 131 (2018), 1803.10835.

\bibitem{Shao:2022cly}
H.-S. Shao and D.~d'Enterria,
\newblock (2022), 2207.03012.

\bibitem{AFP}
The AFP project in ATLAS, Letter of Intent of the Phase-I Upgrade (ATLAS
  Collab.), {\tt http://cdsweb.cern.ch/record/1402470}.

\bibitem{Tasevsky:2015xya}
ATLAS, M.~Taševský,
\newblock AIP Conf. Proc. {\bf 1654}, 090001 (2015).

\bibitem{CT-PPS}
M.~Albrow {\em et~al.},
\newblock CERN Report No. CERN-LHCC-2014-021. TOTEM-TDR-003. CMS-TDR-13, 2014
  (unpublished).

\bibitem{CMS:2020rzi}
CMS, TOTEM, M.~Albrow {\em et~al.},
\newblock CERN Report No. CMS-PAS-EXO-18-014, TOTEM-NOTE-2020-003, 2020
  (unpublished).

\bibitem{CMS:2022zfd}
CMS, TOTEM, M.~Albrow {\em et~al.},
\newblock CERN Report No. CMS-PAS-EXO-21-007, TOTEM-NOTE-2022-005, 2022
  (unpublished).

\bibitem{Aad:2015bwa}
ATLAS Collaboration, G.~Aad {\em et~al.},
\newblock (2015), 1506.07098.

\bibitem{Aaboud:2016dkv}
ATLAS, M.~Aaboud {\em et~al.},
\newblock Phys. Rev. D {\bf 94}, 032011 (2016), 1607.03745.

\bibitem{Aaboud:2017oiq}
ATLAS, M.~Aaboud {\em et~al.},
\newblock Phys. Lett. B {\bf 777}, 303 (2018), 1708.04053.

\bibitem{Chatrchyan:2011ci}
CMS, S.~Chatrchyan {\em et~al.},
\newblock JHEP {\bf 1201}, 052 (2012), 1111.5536.

\bibitem{CMS:2013hdf}
CMS, S.~Chatrchyan {\em et~al.},
\newblock JHEP {\bf 07}, 116 (2013), 1305.5596.

\bibitem{Khachatryan:2016mud}
CMS, V.~Khachatryan {\em et~al.},
\newblock JHEP {\bf 08}, 119 (2016), 1604.04464.

\bibitem{Cms:2018het}
CMS, TOTEM, A.~M. Sirunyan {\em et~al.},
\newblock JHEP {\bf 07}, 153 (2018), 1803.04496.

\bibitem{ATLAS:2020mve}
ATLAS, G.~Aad {\em et~al.},
\newblock Phys. Rev. Lett. {\bf 125}, 261801 (2020), 2009.14537.

\bibitem{Harland-Lang:2019eai}
L.~A. Harland-Lang,
\newblock JHEP {\bf 03}, 128 (2020), 1910.10178.

\bibitem{Harland-Lang:2021zvr}
L.~A. Harland-Lang,
\newblock Phys. Rev. D {\bf 104}, 073002 (2021), 2101.04127.

\bibitem{Harland-Lang:2020veo}
L.~A. Harland-Lang, M.~Tasevsky, V.~Khoze, and M.~Ryskin,
\newblock Eur. Phys. J. C {\bf 80}, 925 (2020), 2007.12704.

\bibitem{Bailey:2022wqy}
S.~Bailey and L.~A. Harland-Lang,
\newblock Phys. Rev. D {\bf 105}, 093010 (2022), 2201.08403.

\bibitem{SuperCHIC}
The SuperCHIC code and documentation are available at {\tt
  http://projects.hepforge.org/superchic/}.

\bibitem{Sjostrand:2014zea}
T.~Sj\"ostrand {\em et~al.},
\newblock Comput. Phys. Commun. {\bf 191}, 159 (2015), 1410.3012.

\bibitem{ATLAS:2017ayi}
ATLAS, M.~Aaboud {\em et~al.},
\newblock Phys. Lett. B {\bf 775}, 105 (2017), 1707.04147.

\bibitem{ATLAS:2021uiz}
ATLAS, G.~Aad {\em et~al.},
\newblock Phys. Lett. B {\bf 822}, 136651 (2021), 2102.13405.

\bibitem{Harland-Lang:2014lxa}
L.~A. Harland-Lang, V.~A. Khoze, M.~G. Ryskin, and W.~Stirling,
\newblock Int.J.Mod.Phys. {\bf A29}, 1430031 (2014), 1405.0018.

\bibitem{Harland-Lang:2015eqa}
L.~A. Harland-Lang, V.~A. Khoze, and M.~G. Ryskin,
\newblock Int.J.Mod.Phys. {\bf A30}, 1542013 (2015).

\bibitem{Kaidalov:2003fw}
A.~B. Kaidalov, V.~A. Khoze, A.~D. Martin, and M.~G. Ryskin,
\newblock Eur.Phys.J. {\bf C31}, 387 (2003), hep-ph/0307064.

\bibitem{Harland-Lang:2010ajr}
L.~A. Harland-Lang, V.~A. Khoze, M.~G. Ryskin, and W.~J. Stirling,
\newblock Eur. Phys. J. C {\bf 69}, 179 (2010), 1005.0695.

\bibitem{Harland-Lang:2018hmi}
L.~A. Harland-Lang, V.~A. Khoze, M.~G. Ryskin, and M.~Tasevsky,
\newblock JHEP {\bf 04}, 010 (2019), 1812.04886.

\bibitem{ATLAS:2014cnc}
ATLAS, G.~Aad {\em et~al.},
\newblock Phys. Rev. D {\bf 90}, 112015 (2014), 1408.7084.

\bibitem{Cerny:2020rvp}
K.~\v{C}ern\'y, T.~S\'ykora, M.~Ta\v{s}evsk\'y, and R.~\v{Z}leb\v{c}\'\i{}k,
\newblock JINST {\bf 16}, P01030 (2021), 2010.00237.

\bibitem{Tasevsky:2014cpa}
M.~Tasevsky,
\newblock Int.J.Mod.Phys. {\bf A29}, 1446012 (2014), 1407.8332.

\bibitem{Jaeckel:2012yz}
J.~Jaeckel, M.~Jankowiak, and M.~Spannowsky,
\newblock Phys. Dark Univ. {\bf 2}, 111 (2013), 1212.3620.

\bibitem{Aad:2012tfa}
ATLAS, G.~Aad {\em et~al.},
\newblock Phys.Lett. {\bf B716}, 1 (2012), 1207.7214.

\bibitem{CMS:2021kom}
CMS, A.~M. Sirunyan {\em et~al.},
\newblock JHEP {\bf 07}, 027 (2021), 2103.06956.

\bibitem{ATLAS:2022fnp}
ATLAS, G.~Aad {\em et~al.},
\newblock (2022), 2202.00487.

\end{thebibliography}
\bibliographystyle{h-physrev}

\end{document}